# Use-It-or-Lose-It Policies for the Available Bit Rate (ABR) Service in ATM Networks[1]

Shivkumar Kalyanaraman, Raj Jain, Rohit Goyal, Sonia Fahmy and Seong-Cheol Kim [2]
Department of Computer and Information Science
The Ohio State University
2015 Neil Ave., Columbus, OH 43210-1277
E-mail: {*shivkuma, jain, goyal, fahmy*}@cis.ohio-state.edu

**Abstract**

The Available Bit Rate (ABR) service has been developed to support 21st century data applications over Asynchronous Transfer Mode (ATM). The ABR service uses a closed-loop rate-based traffic management framework where the network divides left-over bandwidth among contending sources. The ATM Forum traffic management group also incorporated open-loop control capabilities to make the ABR service robust to temporary network failures and source inactivity. An important problem addressed was whether rate allocations of sources should be taken away if sources do not use them. The proposed solutions, popularly known as the Use-It-or-Lose-It (UILI) policies, have had significant impact on the ABR service capabilities. In this paper we discuss the design, development, and the final shape of these policies and their impact on the ABR service. We compare the various alternatives through a performance evaluation.

**Keywords:** Asynchronous Transfer Mode (ATM), Available Bit Rate (ABR), traffic management, congestion control.

# 1 Introduction

The applications of the 21st century are expected to have diverse quality of service (QoS) requirements. High-speed networks are providing multiple classes of service tailored to support such requirements. Of these, classes with higher priority are used by audio, video, and real-time applications, while data applications typically use lower priority classes. Network switches first allocate link bandwidth to higher priority classes and give the left-over band-width to the lower priority classes. As a result, the bandwidth available for data applications is variable. Further, data applications are bursty, i.e., they have active and idle transmission periods and may not always utilize their bandwidth allocations. An important traffic management problem is how to allocate bandwidth among applications which may or may not use their allocations. This problem was debated for over a year in the ATM Forum in the context of traffic management for the Available Bit Rate (ABR) service.

[1]Submitted to IEEE Journal on Selected Areas in Communications (JSAC), January 1997. Available from http://www.cis.ohio-state.edu/~jain/papers.html

[2]Seong-Cheol Kim is with Samsung Electronics Co. Ltd

The ABR service in ATM networks has been developed for applications which expect cell loss guarantees, but can control their data rate dynamically as demanded by the network [1]. The ATM Forum Traffic Management group adapted a rate-based end-to-end framework to allow fair and efficient control of ABR traffic [17, 15]. The main components of the framework are the source end system (SES), the switch, and the destination end system (DES). The ABR source sends data at the Allowed Cell Rate (ACR) which is less than a negotiated Peak Cell Rate (PCR). Immediately after establishing a connection, ACR is set to an Initial Cell Rate (ICR), which is also negotiated with the network. The SES (source) sends an Resource Management (RM) cell after transmitting Nrm-1 cells, where Nrm is a parameter. Among the RM cell fields, the Current Cell Rate (CCR) field informs the network about the source's ACR, and the Explicit Rate (ER) field is used by the network to give its rate feedback. The DES (destination) simply returns RM cells back to the source.

The ABR framework is predominantly closed-loop, i.e., sources normally change their rates in response to network feedback. Another form of control is open-loop control where sources change their rates independent of network feedback. Open-loop control can complement closed-loop control when the network delays are large compared to application traffic chunks, or when network feedback is temporarily disrupted. It is also useful to control applications which are bursty or source-bottlenecked. Bursty application traffic alternates between active periods (application has data to send) and idle periods (application has no data to send). Source-bottlenecked applications cannot sustain a data rate as high as the network allocated rate. The ATM Forum debated on the issue of using open-loop control to reduce rate allocations of sources which do not use them. The proposed solutions, popularly known as the Use-It-or-Lose-It (UILI) policies, have had significant impact on the ABR service capabilities. In this paper, we discuss and evaluate these policies, and their implications on the ABR service.

The paper is organized as follows. Section 2 discusses the issues in the design of UILI policies. We then discuss early UILI proposals in section 3. We identify the problems with the early proposals in section 4 and present the final set of proposals which were debated in the ATM

Forum in section 5. We then evaluate the performance of various alternatives in Section 6 and summarize the implications of UILI on ABR in section 7.

## 2 Issues in Use-It-or-Lose-It

When some VCs' present bursty or source-bottlenecked traffic, the network may experience underload even after rate allocation. It then allocates higher rates to all VCs without first taking back the unused allocations. As a result, the underloading sources retain their high allocations without using them. When these sources suddenly use their allocations, they overload the network.This problem is called "ACR Retention." A related problem is "ACR Promotion" where a source intentionally refrains from using its allocation aiming to get higher allocations in later cycles. The effect of ACR Retention/Promotion is shown in Figure 1. In the figure, before time $t_0$ the source rate is much smaller than its ACR allocation. The ACR allocation remains constant. At time $t_0$, the source rate rises to ACR and the network queues correspondingly rise. These problems were first identified by Barnhart [2].

A solution to this problem is to detect an idle or source-bottlenecked source and reduce its rate allocation before it can overload the network. But this has an important side effect on bursty sources. If the rates are reduced after every idle period and the active periods are short, the aggregate throughput experienced by the source is low. This tradeoff was discovered and studied carefully in the ATM Forum. The solutions proposed are popularly known as the Use-It-or-Lose-It (UILI) policies, referring to the fact that the source's ACR is reduced (lost) if it is not used.

## 3 Early UILI Proposals

The UILI function can be implemented at the SES (source-based) or at the switch (switch-based) or at both places. The early UILI proposals were all source-based. In these proposals, the test for ACR retention is done when an RM cell is being sent by the source. If ACR retention is detected, the source's ACR is immediately reduced using a rate reduction algorithm. Further, to prevent network feedback from overriding the ACR reduction, some proposals ignore the next feedback from the switch (if the feedback requests a rate increase). Over the February, April, May and June 1995 meetings of the ATM Forum, several UILI proposals were considered. The proposals differ in how the ACR retention is detected (additive or multiplicative metric), and in the algorithm used to reduce ACR.

In February 1995, Barnhart proposed a formula which reduced ACR as a function of the time since the last RM cell was sent or rate decrease was last done:

$$\text{ACRn} = \text{ACRo}(1 - \text{T} \times \text{ACRo}/\text{RDF})$$

ACRn is the new ACR and ACRo is the old ACR. The time 'T' in the formula is the time which has transpired since the last *backward* RM cell was received or since the last ACR decrease. RDF is the rate decrease factor which is normally used to calculate the new rate for single-bit feedback. However, it is reused in the reduction formula to avoid choosing a new parameter. ACR retention is detected when the source has sent out $k$ RM cells ($k$ is the Time out Factor (TOF) parameter) but does not hear from the network or has not decreased its rate during the same period.

In April 1995, several flaws with this proposal were corrected. Further, the ACR decrease function was found to be too aggressive and was changed to a harmonic function:

$$1/\text{ACRn} = 1/\text{ACRo} + \text{T}/\text{RDF}$$

The time 'T' in the function is now the time which has transpired since the last *forward* RM cell was sent. In the May and June 1995 meetings several other side effects were identified and corrected. For example, it was felt that the decrease function should not reduce the ACR below the negotiated Initial Cell Rate (ICR), because the source is allowed to start at that rate after an idle period. Kenney [5] observed that the harmonic ACR reduction formula was difficult to implement and proposed a linear reduction formula, which was similar to, but less aggressive than the February proposal:

$$\text{ACRn} = \text{ACRo}(1 - \text{T} \times \text{TDF})$$

'TDF' is a new parameter called "Timeout Decrease Factor". Incorporating these changes, the ABR SES (source) specification in August 1995 read as follows:

*"5. Before sending a forward in-rate RM-cell, if the time T that has elapsed since the last in-rate forward RM-cell was sent is greater than TOF*Nrm cell intervals of (1/ACR), and if ACR > ICR, then:*
*a) ACR shall be reduced by at least ACR * T * TDF, unless that reduction would result in a rate below ICR, in which case ACR shall be set to ICR, and TDF is equal to TDFF/RDF times the smallest power of 2 greater or equal to PCR,* TDFF = $\{\ 0, 2^i, 2^j, 2^l\ \}$ *(2 bits), where the values of the integers* $i, j,$ *and* $l$ *are to be determined in the specification.*
*b) ACR shall not be increased upon reception of the next backward RM-cell."*

The above UILI rule will also be interchangeably called "rule 5" henceforth, referring to the rule number in the ABR SES specification. The two parts are called “rule 5a” and “rule 5b” respectively.

# 4 Problems and Side Effects of Early Proposals

In August 1995, Anna Charny et al [6] pointed out certain undesirable side effects in the above proposal. In particular, sources experience performance degradation in the transient phase when they increase from low ACR to high ACR. As a result, the links may be underutilized for a long period of time.

## 4.1 Worst Case Performance

The worst case occurs when ICR is small and the source rises to a high rate from a very low rate, and when the backward RM cell (BRM) is received just before a forward RM cell (FRM) is sent. The BRM carries the network feedback and asks the source to increase its rate to a value greater than TOF $\times$ (old rate). When the FRM is sent, the measured source rate S is close to the earlier low rate. This results in triggering UILI and the reduction of ACR by ACR $\times$ T $\times$ TDF. Now ACR is large and T is also large since it depends on the earlier low rate. Hence, ACR is reduced by a large amount upto ICR. Since ICR again is a small value, the cycle repeats when the BRM is received just before a FRM is sent. As a result, a source starting from a low ICR may never send at a rate higher than ICR.

## 4.2 Bursty and RPC Traffic Performance

Charny et al [6] also observed that bursty traffic having low ICR experienced a long-term performance degradation due to UILI resulting in large ACR fluctuations. Further, rule 5b prevents the increase of the source rate even though the network may have bandwidth available. In such bursty traffic configurations, it was found that rule 5a without rule 5b yielded better performance than both the parts together. However there was no way to selectively turn off rule 5b. Hence, it was decided to introduce a PNI (Prohibit No Increase) bit which when set turns off rule 5b selectively. Note that this also allows us to turn off rule 5 completely if TDF is also set to zero.

The performance degradation due to remote procedure call (RPC) ping-pong type traffic was independently observed by Bennet et al [7]. These authors pointed out that such applications may not want their rates to be decreased or reset to ICR after every idle period. They also suggested that UILI be performed by the switch and the source-based UILI be left optional.

We note that these side effects of rule 5 are not seen when the source is in the steady state (with source rate approximately equal to ACR) or in the transient phase when the source is decreasing from a high ACR to a low ACR. The main problem seemed to be due to the fact that the decrease function was proportional to T resulting in large ACR decreases after an ACR increase, leading to ramp-up delays.

Another problem which emerged was that some parameters like RDF and ICR were being used in multiple rules. Hence, choosing optimal values for these parameters became difficult due to their various side effects. These problems were addressed in the new set of proposals in December 1995 when the issue was voted upon to arrive at a final decision.

# 5 December 1995 Proposals

There were three main proposals in December 1995: the time-based proposal [8, 9, 10], our count-based proposal [11], and the switch-based proposal [12]. The time-based and the count-based proposals were later combined into one joint proposal. The ATM Forum voted between the switch-based proposal and the joint source-based proposal.

## 5.1 Unresolved UILI Issues

The following were the unresolved issues in UILI in December 1995. Essentially, a UILI proposal which works for both source-bottlenecked and bursty sources was desired.

- How to avoid UILI from affecting the normal rate increase (ramp up) of sources ?
- How long should the switch feedback be ignored after an ACR adjustment ?
- How to ensure good throughput and response time for bursty sources having small, medium and large active periods, when the idle periods are small, medium or large ?
- The floor of the August 1995 UILI ACR reduction function is ICR. If the source rate, S, is larger than ICR, the ACR may be reduced below the source rate down to ICR. We want a reduction function which does not decrease the ACR below the source's rate, S.

- "Headroom" measures how much the ACR is greater than the source rate, S, when it is declared as not an ACR retaining source. Should the headroom be multiplicative (ACR $\leq$ TOF $\times$ S) or additive (ACR $\leq$ S + headroom) ? Is a separate headroom parameter necessary (to avoid depending on ICR) ?
- Can UILI be done effectively in the switch ?
- Under what circumstances is UILI unnecessary or harmful ?

## 5.2 Count-Based UILI Proposal

The count-based UILI proposal [11] was made by us. It solved a large subset of the above problems and presented results of an extensive study on bursty traffic behavior.

### 5.2.1 Count vs Time

First, the count-based proposal removes the dependency of the ACR reduction function on the time factor, T, which is the time since the last FRM is sent. The reduction formula suggested is:

$$\text{ACR} = \text{ACR} - \text{ACR} \times \text{TDF}$$

The proposal is called "count-based" because a constant ACR decrease is achieved by triggering UILI $n$ times. On the other hand, the time-based UILI decreases the ACR proportional to the time factor, T.

### 5.2.2 Multiplicative vs Additive Headroom

The count-based proposal uses an additive headroom for ACR detection (ACR $\leq$ S + headroom). Recall that if the ACR of the source is within the headroom, UILI is not triggered. The problem with multiplicative headroom (ACR $\leq$ TOF $\times$ S) used in the August 1995 proposal is that depending upon the value of S it results in a large difference between ACR and source rate, S. A large difference (ACR - S) results in large network queues when the source suddenly uses its ACR.

The additive headroom allows only a constant difference (ACR $-$ S) regardless of the source rate, S. The queue growth is hence bounded by a constant: $(ACR - S)\times$ Feedback Delay $\times$ Number of Sources. Hence, the additive headroom provides better network protection than the multiplicative headroom. The difference between the multiplicative and additive headroom is shown in Figure 2. Further, the latter is easier to implement since fewer multiply operations are required.

### 5.2.3 Floor of the ACR Reduction Function

We also observed that the floor of the August 1995 UILI ACR reduction function is ICR and independent of the source rate, S. This is problematic because if S is larger than ICR, the ACR may be reduced below the source rate down to ICR. Therefore, we use a different floor function (S + headroom) which ensures that the ACR does not decrease below S or the headroom. This floor function ensures that if the headroom equals the ICR, the ACR is guaranteed not to decreased below ICR.

### 5.2.4 Normal Rate Increase (Ramp Up)

The August 1995 proposal inhibited the ACR ramp up from a low rate because it triggered UILI immediately after the rate increase. Further, the amount of decrease could be large as explained in section 4.1.

Though our proposal does trigger UILI after ramp up from a low rate, it only reduces ACR by a step $\Delta$ = ACR $\times$ TDF. The next BRM cell brings the rate back to the ACR value before the decrease. If TDF is small, UILI is no longer triggered. For larger values of TDF, UILI may still be triggered multiple times. But, our new floor function ensures that the source rate consistently increases by at least the "headroom" value and eventually UILI is no longer triggered.

The count-based proposal also demonstrates a technique which avoids all oscillations due to normal rate increase. The UILI test is disabled *exactly once* after a normal rate increase. This allows the source rate to stabilize to the new (high) rate before the next UILI test, and thus UILI is not unnecessarily triggered. We use a bit called the PR5 ("Prohibit Rule 5") bit which is enabled whenever there is a normal rate increase. The bit is cleared otherwise.

This technique also has one important side effect. Consider a source which is using its ACR allocation but suddenly becomes idle. Using the RM cells remaining in the network, the network may request a rate increase during the idle period. According to the above technique, the UILI test is disabled exactly once when the source becomes active again. Now observe that the first FRM cell opportunity after an idle period is the only opportunity for the source to reduce its ACR using UILI. This is because the memory of the prior idle period is lost when the next FRM is sent. As a result, UILI is never triggered. However, the PR5 technique is not necessary and can be disabled if TDF is chosen to be small.

Table 1: BRM Actions In The Different Regions Of Count-Based UILI

| Region | Trigger UILI | Increase On BRM | Decrease On BRM |
|---|---|---|---|
| A | Yes unless PR5 | No | Yes |
| B | No | No | Yes |
| C | No | Yes | Yes |
| D | No | Yes | Yes |

#### 5.2.5 Action on BRM

We observed that the ACR reduction function alone is not enough to ensure that that ACR retention is eliminated. For example, the August 1995 proposal requires that if the immediately next BRM feedback, after an UILI ACR reduction, requests a rate increase, and the PNI bit is not set, the BRM feedback is ignored. However, subsequent feedbacks may undo the ACR reduction and the problem of ACR retention still persists.

The count-based proposal ignores the BRM feedback as long as the source does not use its ACR allocation. The proposal uses the headroom area as a hysterisis zone in which network feedback to increase ACR is ignored. The proposal defines four regions of operation A, B, C, and D, as shown in Figure 3. Region A is called the ACR retention region. In this region, $ACR > SR + Headroom$, and UILI is triggered unless the PR5 bit (if used) is set. Region B is the headroom area. In this region, $ACR \leq SR + Headroom$, but $ACR > SR$. In this region BRM feedback requesting increase is ignored. Region C has the source rate equal to ACR. Region D has source rate greater than ACR. Region D is touched briefly when the ACR decreases and the measured source rate is a mixture of the old and new ACRs. In regions C and D, the source obeys the feedback of the network to increase or decrease its ACR. In these regions, the source is not ACR retaining because its source rate is at least equal to its current ACR allocation. The actions in various regions are shown in Table 1. Note that there is no need for the PNI parameter, since UILI can be disabled by simply setting the parameter TDF to zero.

### 5.2.6 Parameter Selection

The count-based proposal has two parameters: "headroom" and "TDF". We recommended a separate "headroom" parameter is to avoid overloading the ICR parameter. This allows the ICR parameter to be set to a high value based on short-term congestion information. The headroom parameter can be set to a more conservative value. It controls how much the sources can lie about their rates at any time and determines how many cells the switch receives at once. However, as discussed in the simulation results of bursty sources (Section 6.2), very small headroom is not desirable. A value of 10 Mbps is recommended. This allows LANE traffic to go at full Ethernet speed. Smaller values can be used for WANs.

The parameter TDF determines the speed of convergence to the desired UILI goals (region B in Figure 3). Hence, it determines the duration for which the network is susceptible to load due to sources suddenly using their ACRs. Larger values of TDF give faster convergence. However, a low value is preferred for bursty sources as discussed in Section 6.2, and TDF set to zero disables UILI. A value of 1/8 or 1/16 is recommended.

### 5.2.7 Pseudo Code For the Count-Based Proposal

In the pseudo code for the count-based proposal given below, the variable 'ACR_ok' indicates that the source has used its allocated ACR, and is allowed to increase its rate as directed by network feedback. The variable 'PR5' when set conveys the fact that the network has just directed an increase. 'SR' is a temporary variable and is not stored between successive execution of the code. Further, the proposal requested a separate parameter 'headroom' instead of using ICR in the UILI formula.

- **At FRM Send event:**

```
SR = Nrm/T;
ACR_ok = ((ACR ≤ SR) OR (TDF == 0.0));
IF (PR5 == FALSE)
        IF (ACR > SR + headroom)
        ACR = Max(SR + headroom, ACR × (1.0 − TDF));
        ENDIF
ELSE PR5 = FALSE;
```

- **At BRM Receive event:**

```
IF (NI = 0 AND ACR_ok)
```

IF (ACR < ER) PR5 = TRUE ELSE PR5 = FALSE;
ACR = Min(ACR + AIR × PCR, PCR);
ENDIF
ACR = Min(ACR, ER);
ACR = Max(ACR, MCR);

- **Initialization**
  ACR_ok = True;
  PR5 = False;

Note that the comparison (ACR $\leq$ SR) may always yield false due to the fact that cells may be scheduled only at certain fixed slots. There is typically a minimum granularity $\Delta$ which dictates the cell scheduler at the source. To account for this scheduler, the comparison may be replaced by (ACR $\leq$ SR + $\Delta$).

### 5.3 Time-Based UILI Proposal

The time-based UILI proposal has a ACR reduction function which depends upon the time T since the last FRM was sent. While this aspect is similar to the August 1995 UILI proposal, the other changes suggested are:

1. The time-based proposal also independently observes the problem with using ICR as the floor of the reduction function (as discussed in Section 5.2.3). The proposal suggests two possible floor values:

   a) $ACR_{max}$ = Max(ICR, TOF × SR)
   b) $ACR_{max}$ = ICR + SR

2. IF ( ACR > $ACR_{max}$ )
   $ACR_{new}$ = Max( ACR × (1 − T/Tc), $ACR_{max}$ );

   The recommended value for Tc is Max(ADDF × FRTT, TBE/PCR), where ADDF has a default value of 2. FRTT is the Fixed Round Trip Time measured at connection setup.

The ACR reduction formula decreases ACR depending upon how long the idle period is compared to the round-trip time. A performance comparison of the count-based and the time-based alternatives is presented in Section 6.

## 5.4 Joint Source-Based UILI Proposal

The time-based and count-based camps agreed on a consensus, which we refer to as the "joint source-based proposal." The proposal uses the count-based reduction function and a constant value for TDF. It uses the new floor of the reduction function and the additive headroom. However, ICR is used in the UILI function instead of the proposed "headroom" parameter. The hysterisis region (region B in Figure 3) suggested by the count-based proposal is not used. Rule 5b remains the same as the August 1995 proposal, and PR5 is not used since TDF is set to a small value (1/16), the count-based reduction formula is used.

The effect of removing the hysterisis region in the joint proposal is shown in Figure 4. In the joint proposal, the source will ignore one ER feedback after reducing the ACR to within the desired threshold. However, it may increase its rate-based upon ER feedback henceforth. The source thus re-enters the danger zone of ACR retention. In the count-based proposal, a source which reaches the desired operating zone (ACR <= SR + ICR), it remains in this region until the source actually uses its ACR allocation.

## 5.5 Switch-Based Proposal

AT&T [12] argued that the UILI function can be implemented in the switches on the following lines:

- Estimate rate of a connection and derive a smoothed average. This requires per-VC accounting at the switches.
- The switch maintains a local allocation for the VC based on the max-min fair allocation and the rate the VC claims to go at, i.e., its CCR.
- Use an "aging" function at the switch which allocates a rate to the VC based on the the ratio of the CCR and the actual rate-estimate. Basically, this function widthdraws the allocations from ACR retaining sources.

  A suggested aging function was ( $e^{\alpha u} - e^{\alpha \delta}$ ) where, $u$ is the ratio of the expected rate and the actual rate, and, $\alpha$ and $\delta$ are parameters. The function has the property that the larger the difference between the CCR and the estimated actual rate, the greater the reduction factor. Essentially, the switch allocates conservatively to sources which it knows are not using their allocations.

A switch-based policy with no support from the source faces problems in handling sources which go idle because idle sources do not send RM cells. The switch may take away the allocation of an idle source after a timeout, but there is no way to convey this information to the idle source, since there are no RM cells from the source. Therefore, the switch-based UILI proposal suggests a simple timeout mechanism at the source which reduces the rate of the source to ICR after a timeout (parameter ATDF) of the order of 500 ms. Note that idle sources which become active before the timeout expires may still overload the network. The proposal does not implement UILI for such sources.

# 6 Simulation Results

In this section, we study the tradeoffs in the UILI design through simulation results. We look at both source-bottlenecked and bursty source configurations and present simulation results for the following five UILI alternatives:

1. No UILI
2. August 1995 UILI proposal
3. Baseline Rule 5 (enhanced August 1995) proposal, where the time-based reduction formula is replaced by the count-based formula, and an additive headroom (equal to ICR) is used in place of the multiplicative headroom.
4. The count-based UILI proposal
5. The time-based UILI proposal

A complete set of simulation results may be found in reference [18].

## 6.1 Source Bottlenecked Configuration

The configuration is a network consisting of five ABR sources (Figure 5) going through two switches to corresponding destinations. All simulation results use ERICA switch algorithm [13]. All links are 155 Mbps and 1000 km long. All VCs are bidirectional, that is, D1, D2, through D5 are also sending traffic to S1, S2 through S5. Some important ABR SES parameter values are given below. The values have been chosen to allow us to study UILI without the effect of other SES rules.

PCR = 155.52 Mbps, MCR = 0 Mbps, ICR = 155.52 Mbps, 1 Mbps
RIF (AIR) = 1, Nrm = 32, Mrm = 2, RDF = 1/512
Crm = Min{TBE/Nrm, PCR × FRTT/Nrm}
TOF = 2, Trm = 100 ms, FRTT = 30 ms, TCR = 10 cells/sec
TBE = 4096 (Rule 6 effectively disabled), CDF (XDF) = 0.5
TDF = {0, 0.125} : {0 ⇒ No rule 5, 0.125 for all versions of rule 5}
PNI = {0, 1} : {1 ⇒ No rule 5b, 0 ⇒ Rule 5b for August 1995 and Baseline UILI}

The simulation is run for 400 ms. For the first half of the simulation (200 ms), all the VCs are source-bottlenecked at 10 Mbps. After t=200 ms, all sources are able to use their allocated rates.

Figure 6 shows the ACR, and the actual source rates for the five UILI alternatives studied. There are six lines in each graph consisting of five ACR values and one actual source rate. Since all five sources are identical, the curves lie on the top of each other. With no UILI implemented (figure 6(a)) the ACR is initially much larger than the actual source rate. At 200 ms, the source rate jumps to the ACR and results in network overload. Figure 6(b) shows oscillatory behavior of the August 1995 proposal due to the wrong floor of the ACR reduction function. The Baseline UILI reaches the goal. However it oscillates between the goal and the network feedback. The count-based UILI converges quickly to the goal and does not have oscillations after reaching the goal. The time-based UILI converges very slowly to the goal. Had the sources started using their ACR allocations earlier (than 200ms), it would have resulted in network overload.

## 6.2 Bursty Sources

Recall that bursty sources have active periods when they send data at the allocated rate and idle periods when they do not have data to send. From the point of view of the bursty application, the following two measures are of interest (figure 7):

- *Burst response time* is the time taken to transmit the burst.
- *Effective throughput* is the average transmission rate of the burst.

Figure 7 shows the arrival and departure of a burst at an end system. The top part of the figure shows a burst which takes a long time to be transmitted, and the bottom part shows one which is transmitted quickly. In the former case, the burst response time is short and effective throughput is higher, and vice versa for the latter case. Note that the effective throughput is related to the size of the burst and the burst response time.

Observe that the UILI goals conflict with the above bursty traffic performance goals. When UILI works, ACR is effectively reduced and a bursty source keeps restarting from low rates after every idle period. This results in a high burst response time which implies reduced performance. We study the effect of the UILI policy for different lengths of the active period: short (burst size is smaller than Nrm), medium (burst time smaller than round trip time (RTT), but burst size larger than Nrm) and large (burst time larger than RTT). Handling the network queues is usually not a problem for short or medium bursts. But it does become important when larger bursts active periods are used. The next section describes a model to generate short, medium and long bursts.

### 6.2.1 Closed-Loop Bursty Traffic Model

We define a new "closed-loop" bursty traffic model as shown in Figure 8. The model consists of cycles of request-response traffic. In each cycle the source sends a set of requests and receives a set of responses from the destination. The next cycle begins after all the responses of the previous cycle have been received and an inter-cycle time has elapsed. There is a gap between successive requests called the inter-request time. The request contains a bunch of cells sent back-to-back by the application at rate PCR and the adapter controls the output rate to ACR.

The model as presented above may roughly represent World Wide Web traffic, transaction-oriented traffic, or client-server traffic. The model is "closed-loop" in the sense that the rate at which cycles (and hence requests) are generated depends upon the responsivity of the network. If the network is congested the response take longer time to come back and the sources do not generate new requests until the previous ones have been responded to. In an "open-loop" traffic model like the packet-train model [16], bursts are generated at a fixed rate regardless of the congestion in the network.

Note that the time between two sets of requests (called a cycle time) is at least the sum of the time to transmit requests, the round-trip time and the inter-cycle time. Thus the idle time between two sets of requests is always greater than the round-trip time. All the RM cells from the previous set of requests return to the source before the new set of requests are sent. When a new burst starts there are no RM cells of the source in the network (ignoring second-order effects).

In our simulations, a cycle consists of one request from the client and one response from the server. We use a small response burst size (16 cells), and vary the request burst size.

### 6.2.2 Single-Client Configuration and Parameter Values

The configuration we use is called the single-client configuration (Figure 9). It consists of a single client which communicates with the server, via a VC which traverses a bottleneck link. An infinite source is used in the background to ensure that the network is always loaded, and any sudden bursts of traffic manifest as queues. All the links run at 155 Mbps.

The response size is kept constant at 16 cells. The request size can be 16, 256 or 8192 for small, meduim or large bursts respectively. The inter-cycle time is chosen to be 1ms. All links are 500km long. The other source parameters are chosen to maximize ACR and disable the effects of other source rules:
ICR = 10 Mbps, TDF = 1/8, TCR = 10 cells/sec
TRM = 100 ms, TBE = 512, CDF = 0 to disable SES Rule 6.

The switch uses the ERICA algorithm [13] to calculate rate feedback. The ERICA algorithm uses two key parameters: target utilization and averaging interval length. The algorithm measures the load and number of active sources over successive averaging intervals and tries to achieve a link utilization equal to the target. The averaging intervals end either after the specified length or after a specified number of cells have been received, whichever happens first. In the simulations reported here, the target utilization is set at 90%, and the averaging interval length defaults to 100 ABR input cells or 1 ms, represented as the tuple (1 ms, 100 cells).

In the following sections, we pictorially describe the simulation results; a full set of graphs may be found in reference [18].

### 6.2.3 Small Bursts

Small bursts are seen in LANE traffic. For example, the ethernet MTU, 1518 bytes is smaller than 32 (Nrm) cells. Since small bursts are smaller than Nrm cells, no RM cells are transmitted during certain bursts. As a result, no SES rules are triggered during these bursts. In other words, the entire burst is transmitted at one rate. However, when RM cells are finally transmitted, UILI is triggered which brings down the ACR to ICR. The source rate, S, is nearly zero due to the short burst time and long idle time. Hence, ICR + S is approximately equal to ICR.

Figure 10 shows the effect of UILI on the source rate of small bursts. The network feedback first arrives when the source is idle, asking it to increase its ACR. The source uses its ACR to almost send the full burst. The first RM cell sent reduces its source rate back to ICR. The source rate goes back to zero when the source is idle. Now, the time-based and count-based proposals differ in the way they respond to subsequent network feedback.

In the time-based proposal, the feedback brought by the next RM cell is ignored because of rule 5b. Now there is no RM cell of the source in the network and at least two bursts are sent at ICR before the next RM cell is sent which results in an ACR increase. Note that the sending of this second RM cell does not decrease the ACR further because ACR is already at ICR. Therefore, on the average one out of every three bursts is sent at a higher rate.

In the count-based proposal, the rate-increase feedbacks are always ignored because the system is in region B (Figure 3). The ACR slowly reduces to ICR and then remains at ICR. Over the long term, all short bursts are sent out at ICR only. This can be improved by using a leaky bucket or GCRA [14] type burst tolerance mechanism where small bursts can be sent at link rate irrespective of ACR or ICR. Other alternatives include choosing a small TDF or a larger ICR. An ICR of 10 Mbps allows LANE traffic (the source of small bursts) to go through at full speed. On the other hand, since the burst is very short, there is not a significant time difference in transmitting the burst at ACR and transmitting it at ICR (assuming ICR is not very small). In such a case, the emphasis then shifts to supporting medium bursts and large bursts efficiently.

#### 6.2.4 Medium Bursts

Medium bursts are expected in ATM backbone traffic or in native mode ATM applications. Medium bursts contain more than Nrm (32) cells, but the active time is shorter than the round trip time. Though multiple RM cells are sent in a single burst, the network feedback for the burst arrives only after the burst has already been transmitted.

As shown in Figure 11, the UILI mechanism triggers once when the first RM cell is sent. In the time-based proposal, the amount of decrease is proportional to the idle time prior to the burst, while in the count-based UILI, the decrease is a constant amount. In the time-based proposal, if the idle time is large, almost the entire burst may be transmitted at ICR. Since, the count-based proposal sends the burst almost at ACR $\times$ (1 $-$ TDF), it provides better burst response. Accordingly, simulation results in reference [18] show that the average source rate experienced by the bursts is higher for the count-based option (120 Mbps) compared to the time-based option (68 Mbps).

### 6.2.5 Large Bursts

Large bursts are expected to be seen in backbone ATM links. Large bursts have a burst time larger than the round trip time. The network feedback returns to the source before the burst completes transmission.

Figure 12 shows the behavior of large bursts with the August 1995 proposal. When the burst starts, UILI triggers when the first RM cell is sent and brings the rate to ICR. Some part of the burst is transmitted at ICR. When network feedback is received, the ACR increases to the network directed value. If ICR is not very low there are no further oscillations and normal increase is not hampered. However if ICR is very low UILI is triggered after the ACR increase bringing the rate down to ACR again. The cycle is repeated and UILI triggers multiple times during the transmission of the burst resulting in low effective throughput and high burst response time.

The time-based UILI avoids the multiple triggering of UILI. It triggers once when the burst starts, and reduces the ACR proportional to the idle time. The count-based UILI also triggers once, and reduces the ACR by a constant value. Since the burst size is large, for large idle times ($>$ RTT), the network protection provided by the count-based technique may be insufficient. However under such conditions a different SES rule (rule 6) can provide the required network protection.

# 7 ATM Forum decision

The ATM Forum debated considerably over the UILI issue in December 1995 before putting the issue to vote. The summary of the arguments were the following:

The UILI policy can be implemented in switches or in NICs (sources) or both. The advantage of switch-only implementation is that NICs are simpler. The advantage of NIC implementation is that switches can be more aggressive in their bandwidth allocation without worrying about long-term implications of any one allocation. Without source-based UILI, the switches have to provision buffers to allow for overallocation of bandwidth.

Finally, the ATM Forum decided not to standardize an elaborate source-based UILI policy. A simple timeout is mandated for the source, where sources keep their rate allocations until a timeout (parameter ATDF, of the order of 500 ms) expires. After the timeout expires, ACR is reduced to ICR. The burden of implementing UILI is on the switches. However, NIC manufacturers can optionally implement a source-based UILI policy. The Informative appendix I.8 of the ATM Traffic Management 4.0 specification [14] briefly describes some source-based policies including the joint source-based proposal. The purpose of this paper has been to describe and evaluate the performance of various options.

# References

[1] R. Jain, "ABR Service on ATM Networks: What is it?," *Network World*, June 24th, 1995.[3]

[2] A.W.Barnhart, "Changes Required to the Specification of Source Behavior," *AF-TM 95-0193*[4], February 1995.

[3] A.W.Barnhart, "Evaluation of Source Behaviour Specification," *AF-TM 95-0267R1*, April 1995.

[4] Lawrence G. Roberts, "RM Return Failure Condition," *AF-TM 95-0340*, April 1995.

[5] John B. Kenney, "Problems and Suggested Solutions in Core Behavior," *AF-TM 95-0564R1*, May 1995.

[6] Anna Charny, Gunter Leeb, Michael Clarke, "Some Observations on Source Behavior 5 of the Traffic Management Specification," *AF-TM 95-0976R1*, August 1995.

[7] Bennett J., Fendick K., Ramakrishnan K.K., Bonomi F., "RPC Behavior as it Relates to Source Behavior 5," *AF-TM 95-0568R1*, May 1995.

[8] A.W.Barnhart, "Evaluation and Proposed Solutions for Source Behavior # 5," *AF-TM 95-1614*, December 1995.

[9] L.G.Roberts, "Operation of Source Behavior # 5," *AF-TM 95-1641*, December 1995.

[3] All our papers and ATM Forum contributions are available through http://www.cis.ohio-state.edu/~jain

[4] Throughout this section, AF-TM refers to ATM Forum Traffic Management sub-working group contributions.

[10] S. Liu, M. Procanik, T. Chen, V.K. Samalam, J. Ormond, "An analysis of source rule # 5," *AF-TM 95-1545*, December 1995.

[11] R. Jain, S. Kalyanaraman, R. Goyal, S. Fahmy, F. Lu, "A Fix for Source End System Rule 5," *AF-TM 95-1660*, December 1995.

[12] K.K. Ramakrishnan, P. P. Mishra and K. W. Fendick, "Examination of Alternative Mechanisms for Use-it-or-Lose-it," *AF-TM 95-1599*, December 1995.

[13] Raj Jain, Shiv Kalyanaraman, Rohit Goyal, Sonia Fahmy, and Ram Viswanathan, "ERICA Switch Algorithm: A Complete Description," *AF-TM 96-1172*, August 1996.

[14] ATM Forum, "ATM Traffic Management Specification Version 4.0," April 1996, available as ftp://ftp.atmforum.com/pub/approved-specs/af-tm-0055.000.ps

[15] R. Jain, "Congestion Control and Traffic Management in ATM Networks: Recent advances and a survey," invited submission to *Computer Networks and ISDN Systems*, October 1996.

[16] R. Jain and S. Routhier, "Packet Trains - Measurement and a new model for computer network trafic," *IEEE Journal of Selected Areas in Communications,* Vol. SAC-4, No. 6, September 1986, pp. 986-995.

[17] Raj Jain, Shiv Kalyanaraman, Rohit Goyal, Sonia Fahmy, "Source Behavior for ATM ABR Traffic Management: An Explanation," *IEEE Communications Magazine*, November 1996.

[18] Shiv Kalyanaraman, Raj Jain, Rohit Goyal, Sonia Fahmy, "A Survey of the Use-It-Or-Lose-It Policies for the ABR Service in ATM Networks," *Technical Report, Dept of CIS, Ohio State University*, 1997.

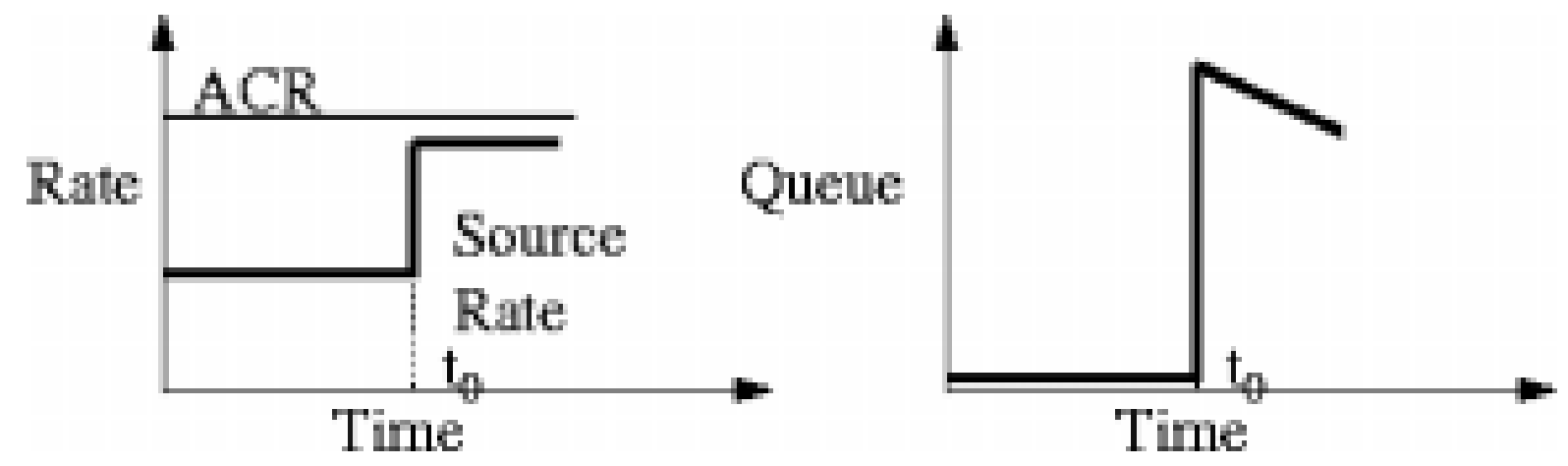

Figure 1: Effect of ACR Retention/Promotion

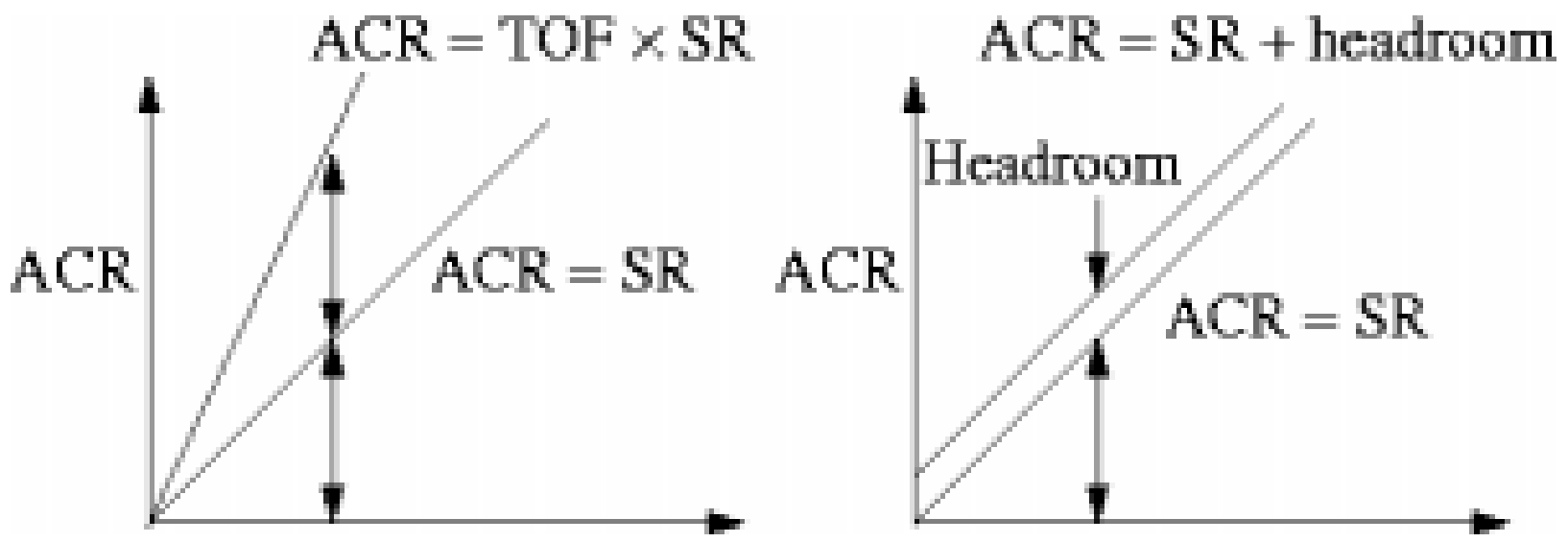

Figure 2: Multiplicative vs Additive Headroom

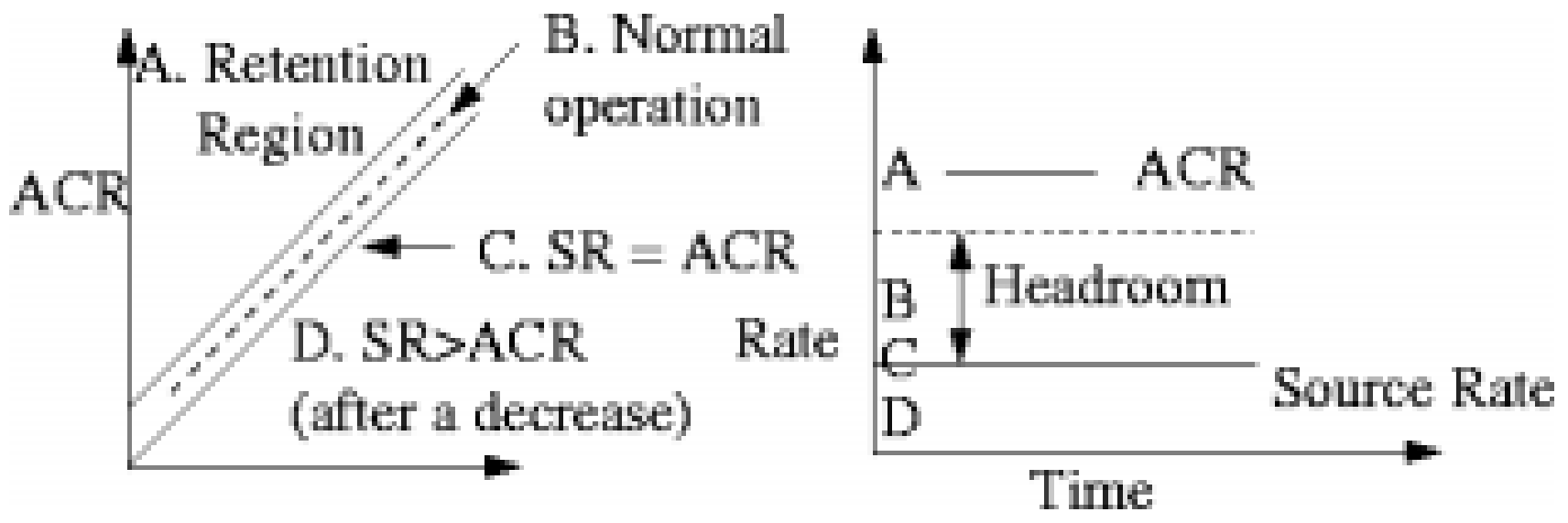

Figure 3: Regions of Operation

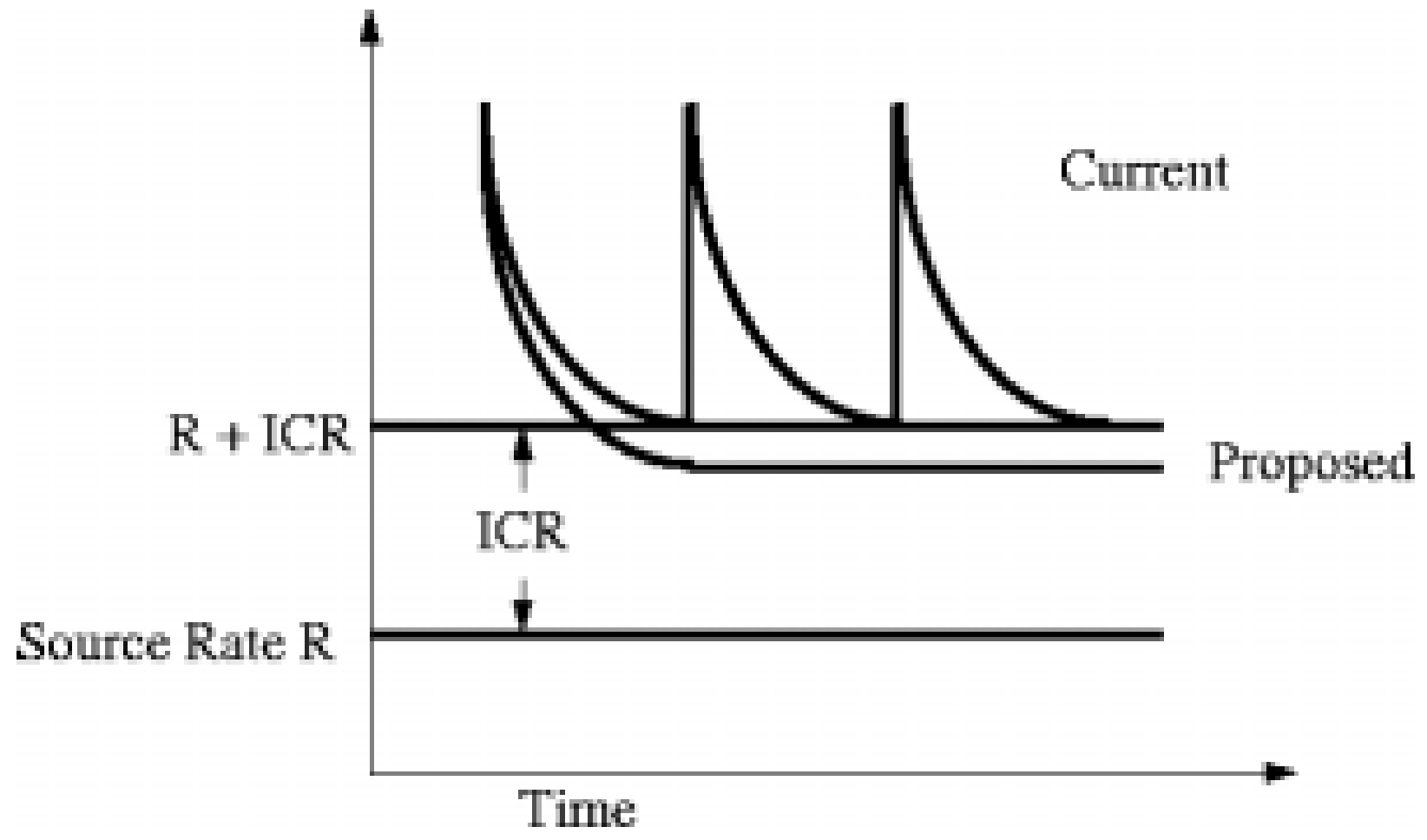

Figure 4: Joint Source-Based UILI Proposal vs Count-Based Proposal

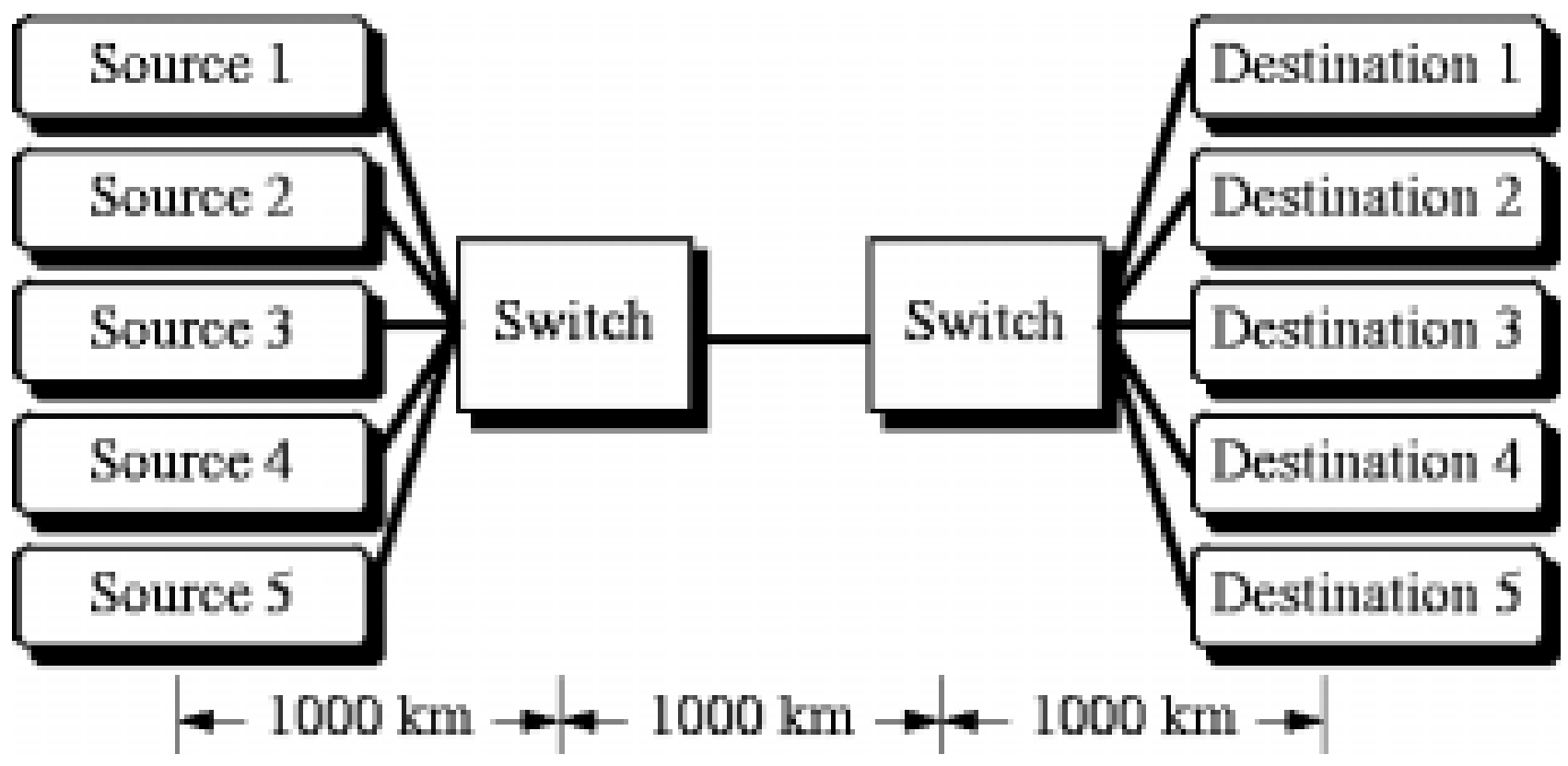

Figure 5: Five Sources Configuration

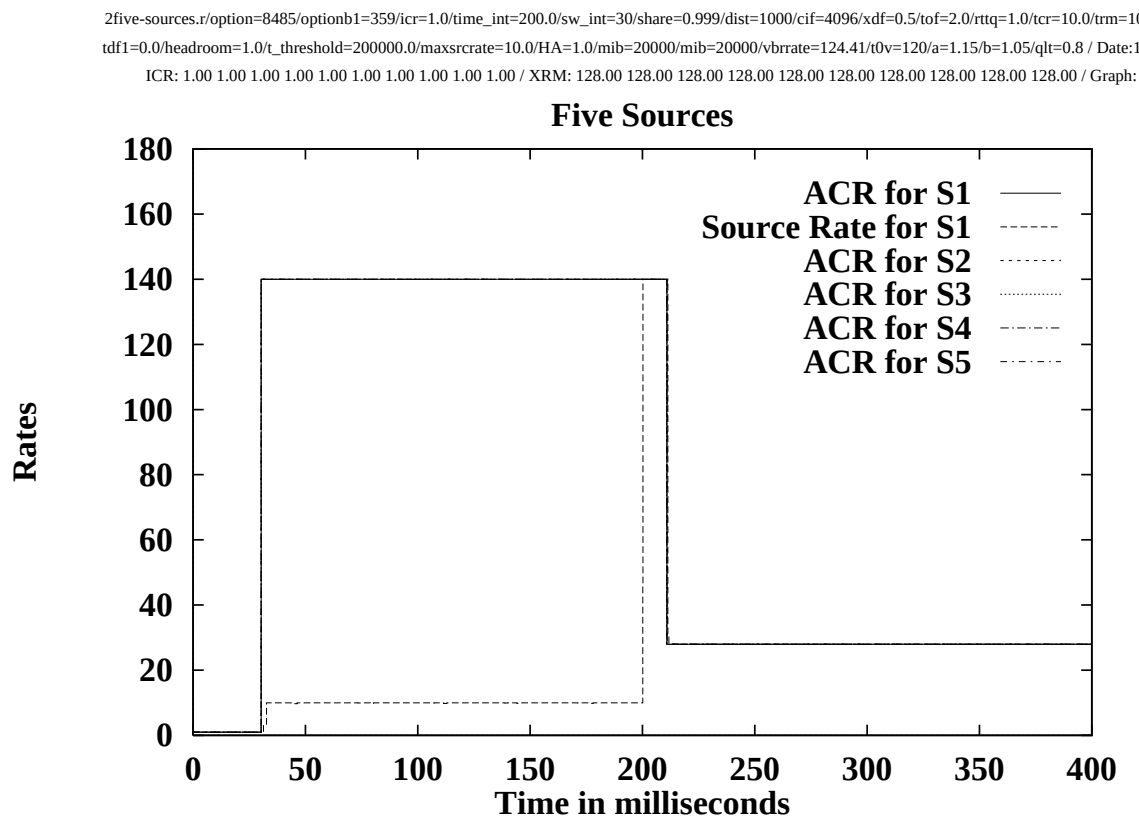

(a) No UILI

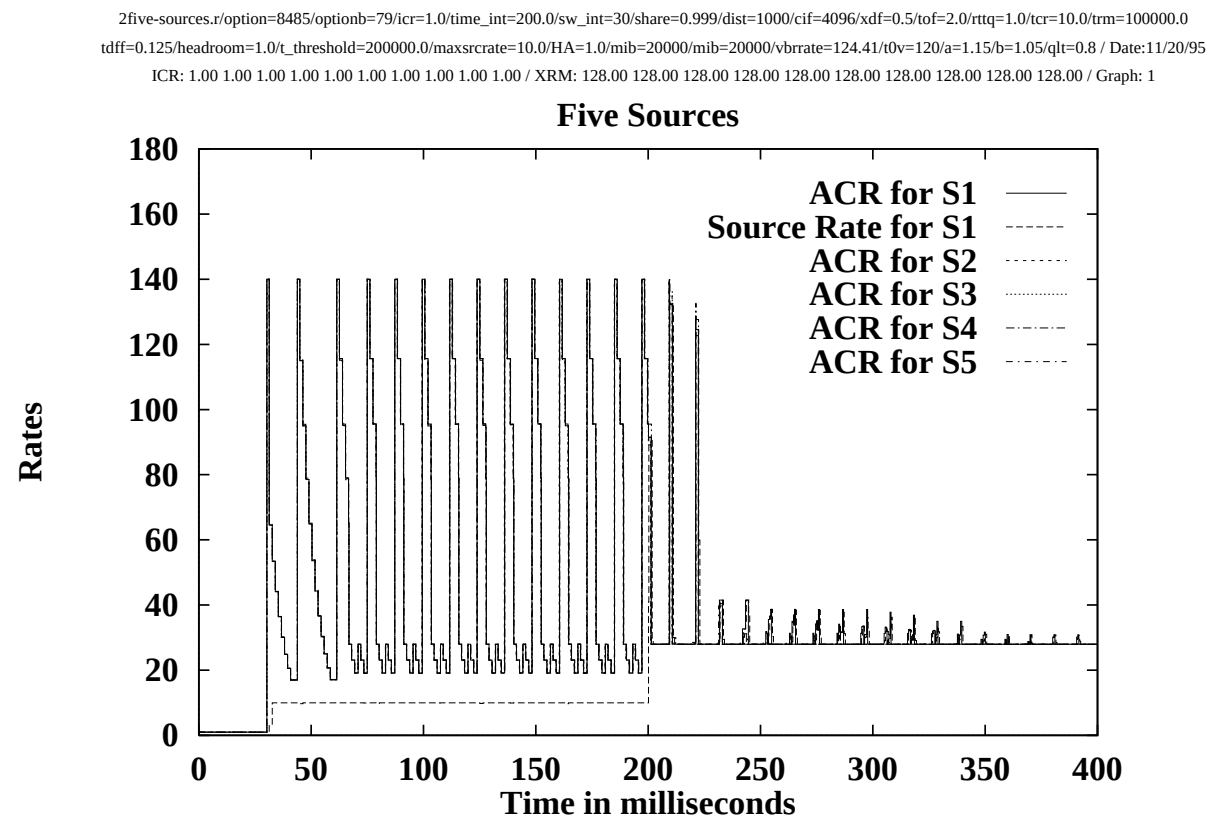

(b) Aug 1995 UILI

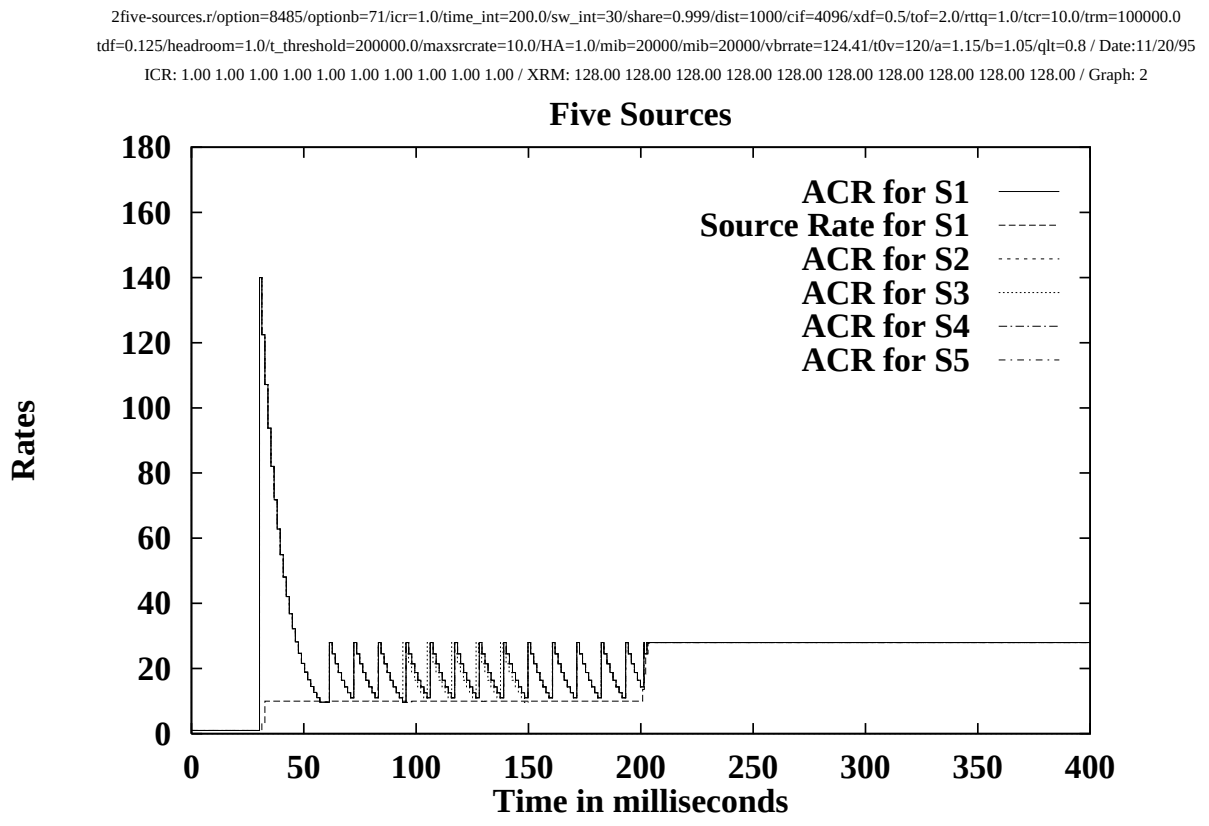

(c) Baseline UILI

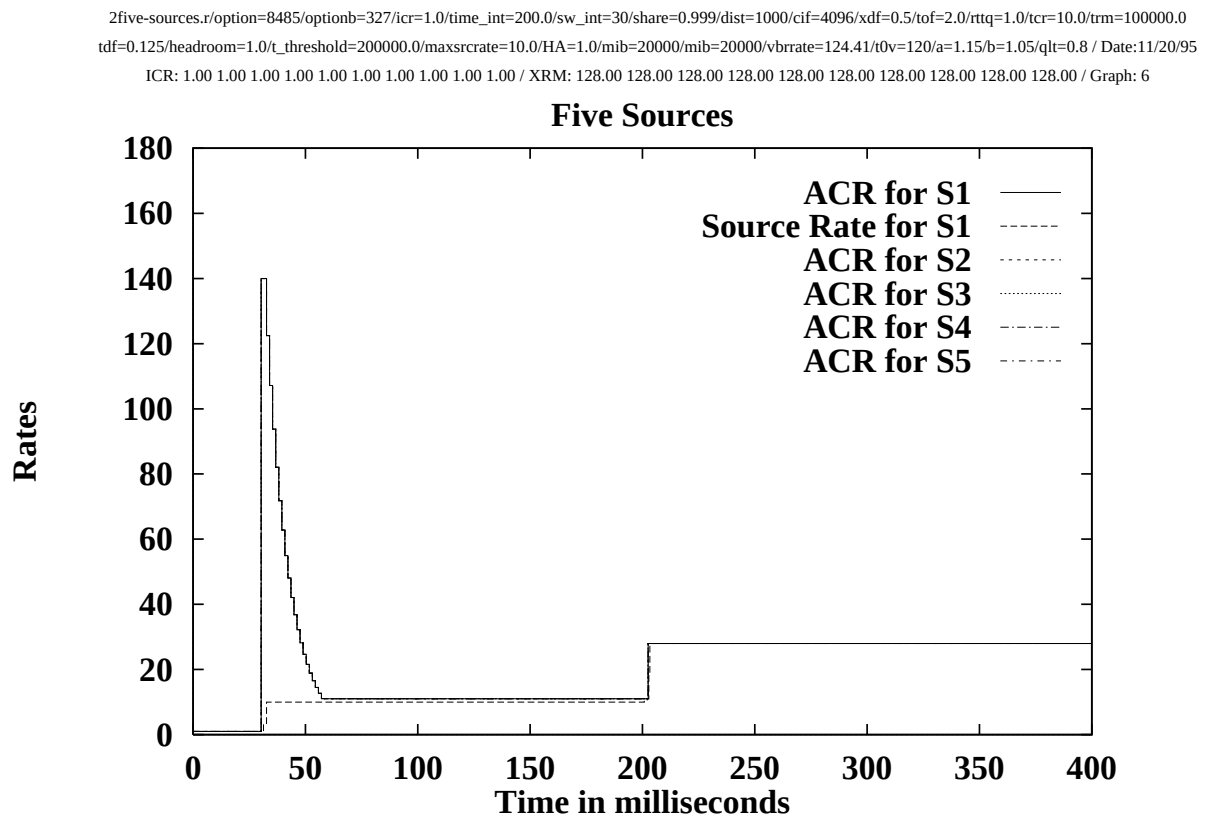

(d) Count-Based UILI

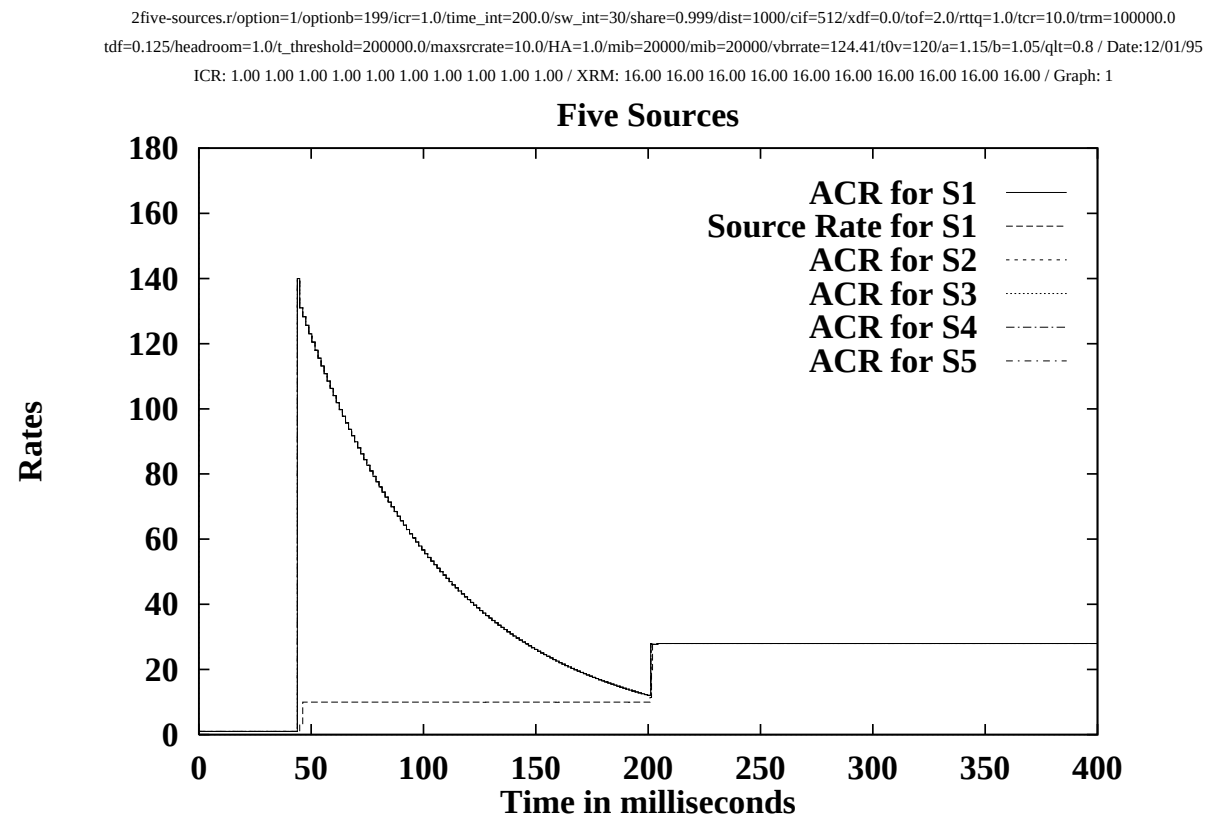

(e) Time-Based UILI

Figure 6: Five Source Config., Rates, ICR=1 Mbps, Headroom=1Mbps, MaxSrcRate=10Mbps for 200ms

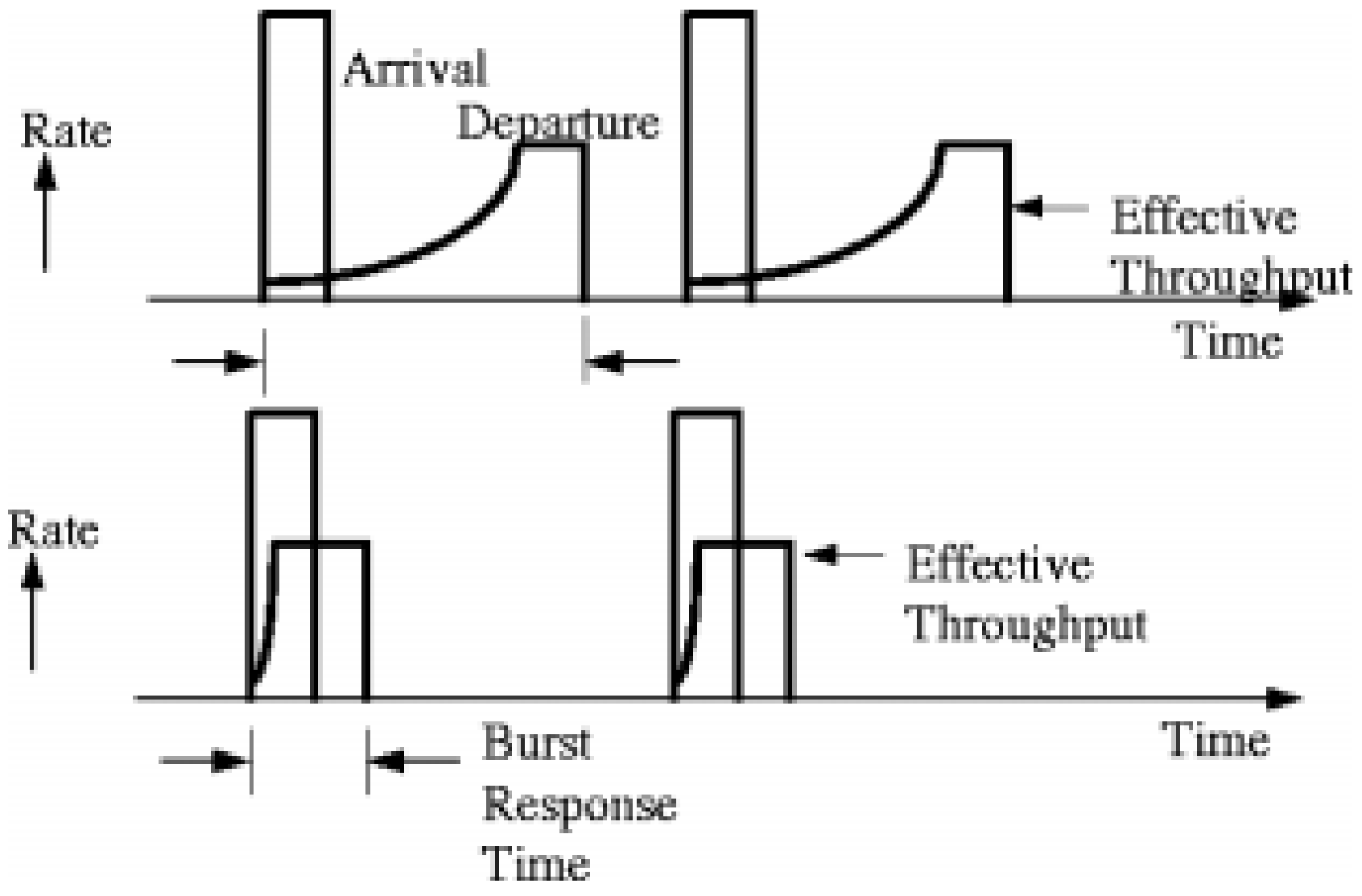

Figure 7: Burst Response Time vs Effective Throughput

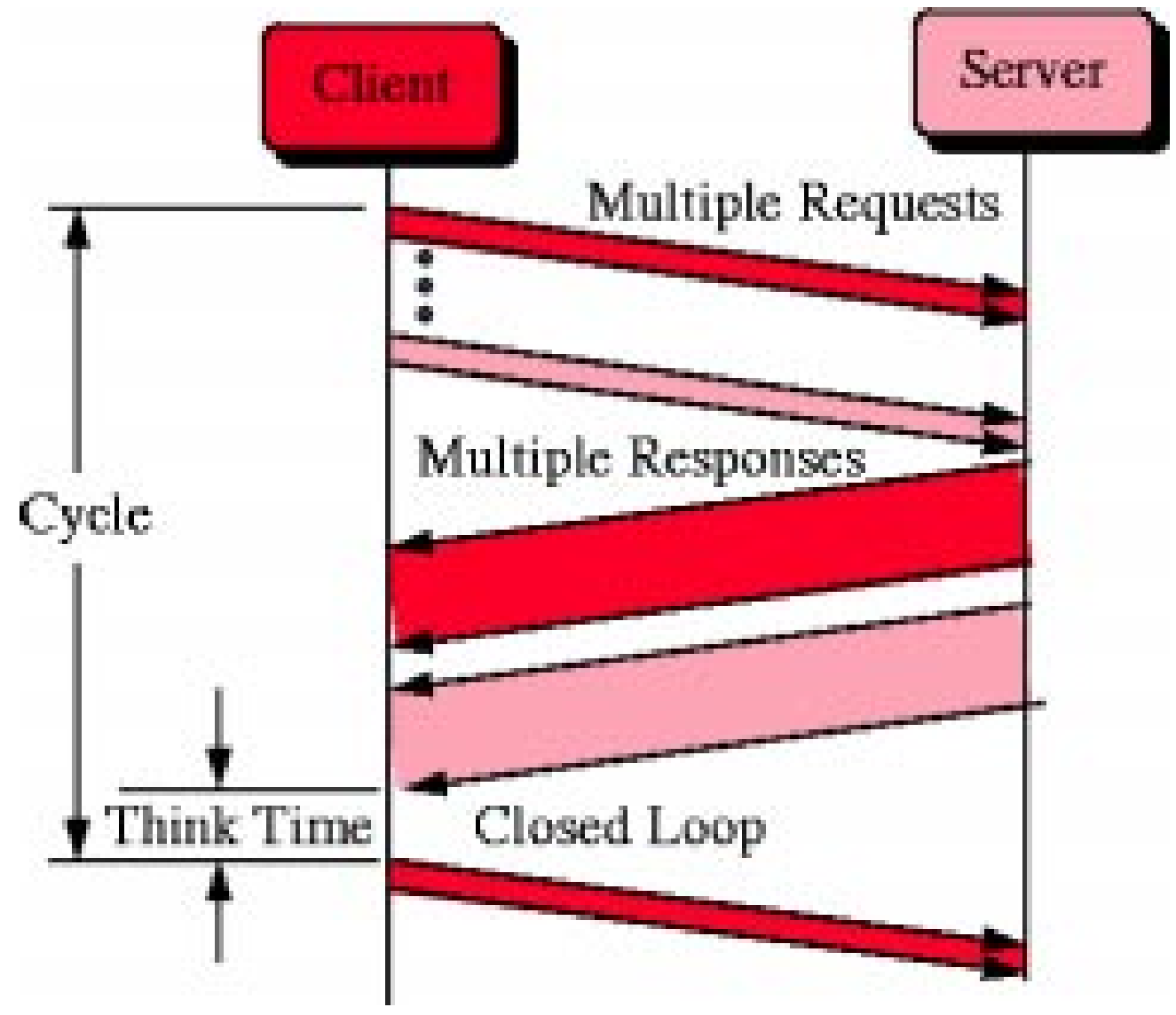

Figure 8: Closed-Loop Bursty Traffic Model

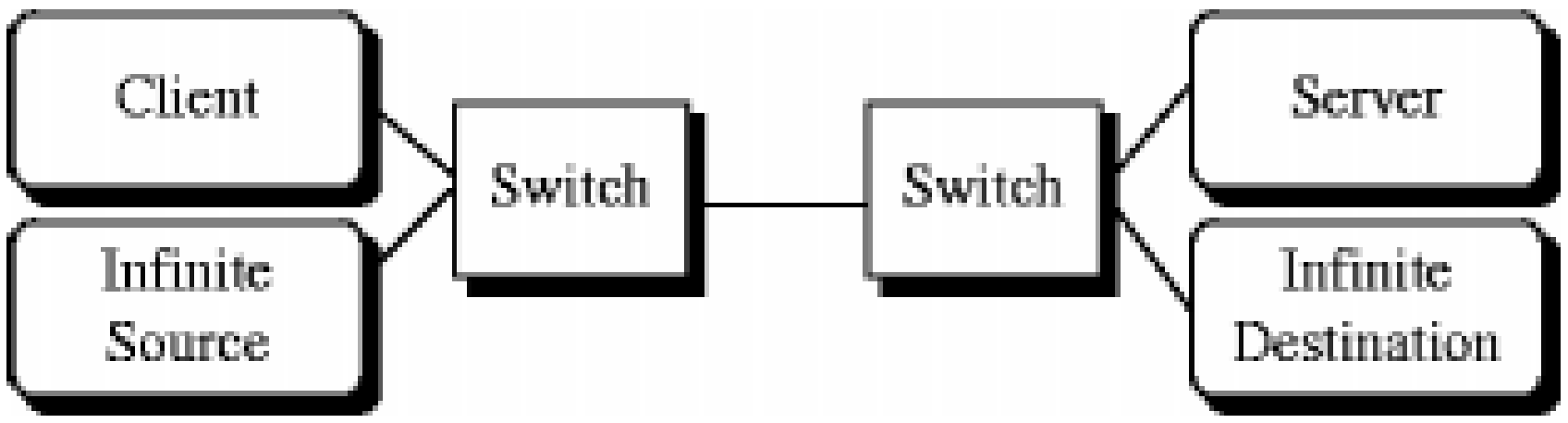

Figure 9: Client-Server Configuration With Infinite Source Background

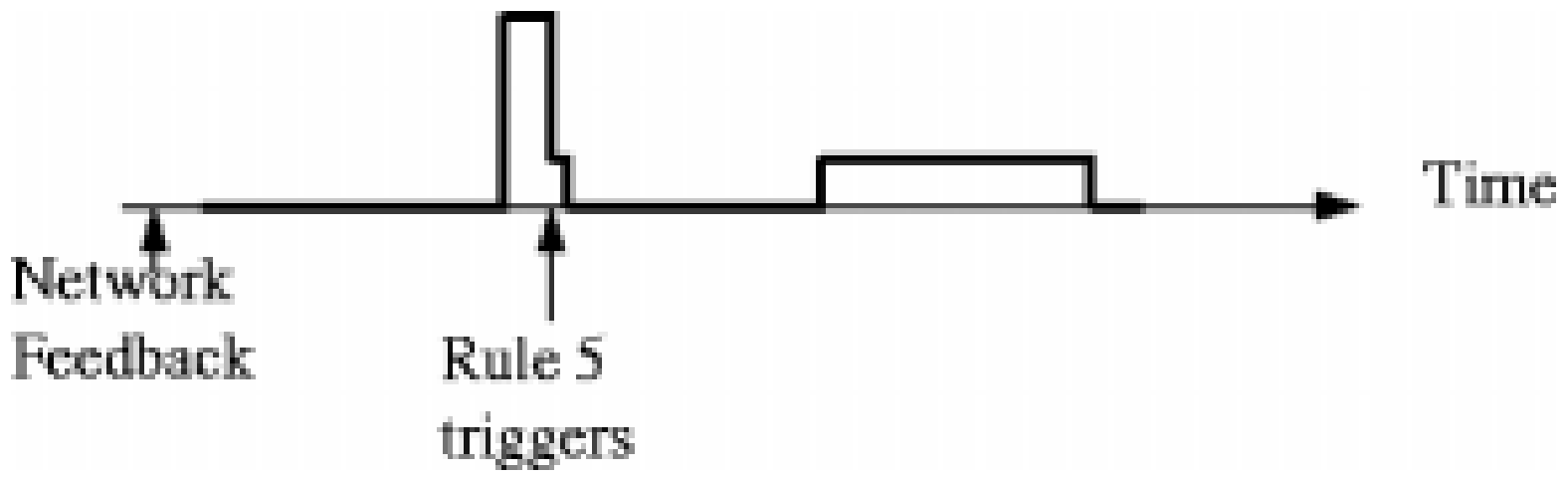

Figure 10: Effect of UILI on Small Bursts

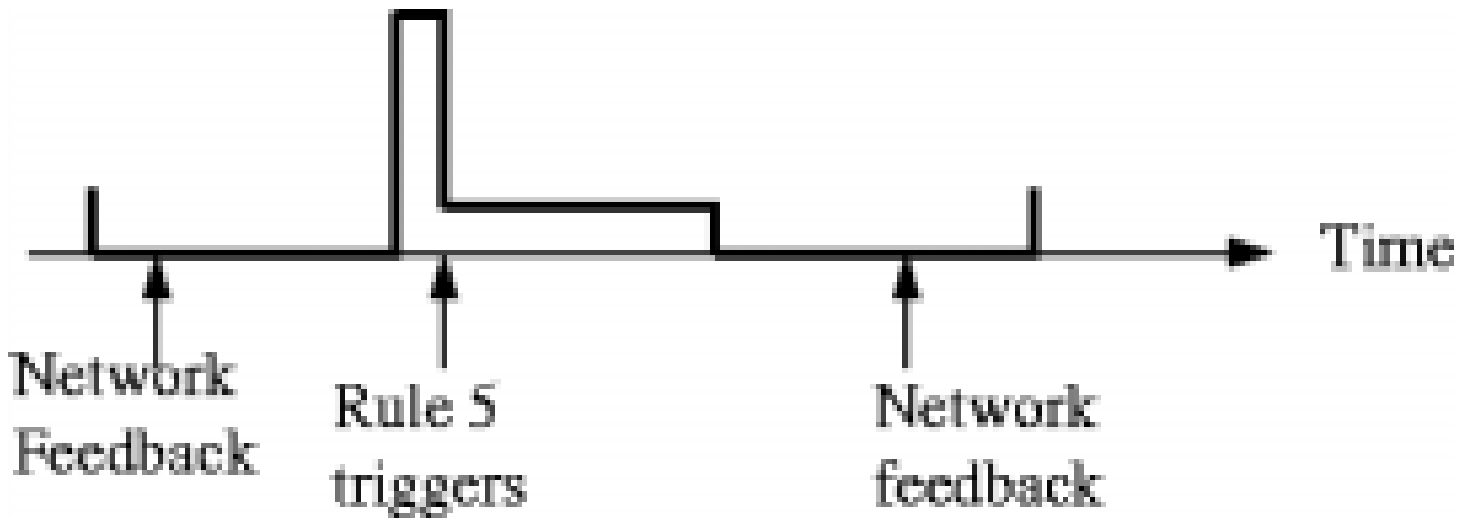

Figure 11: Effect of UILI on Medium Bursts

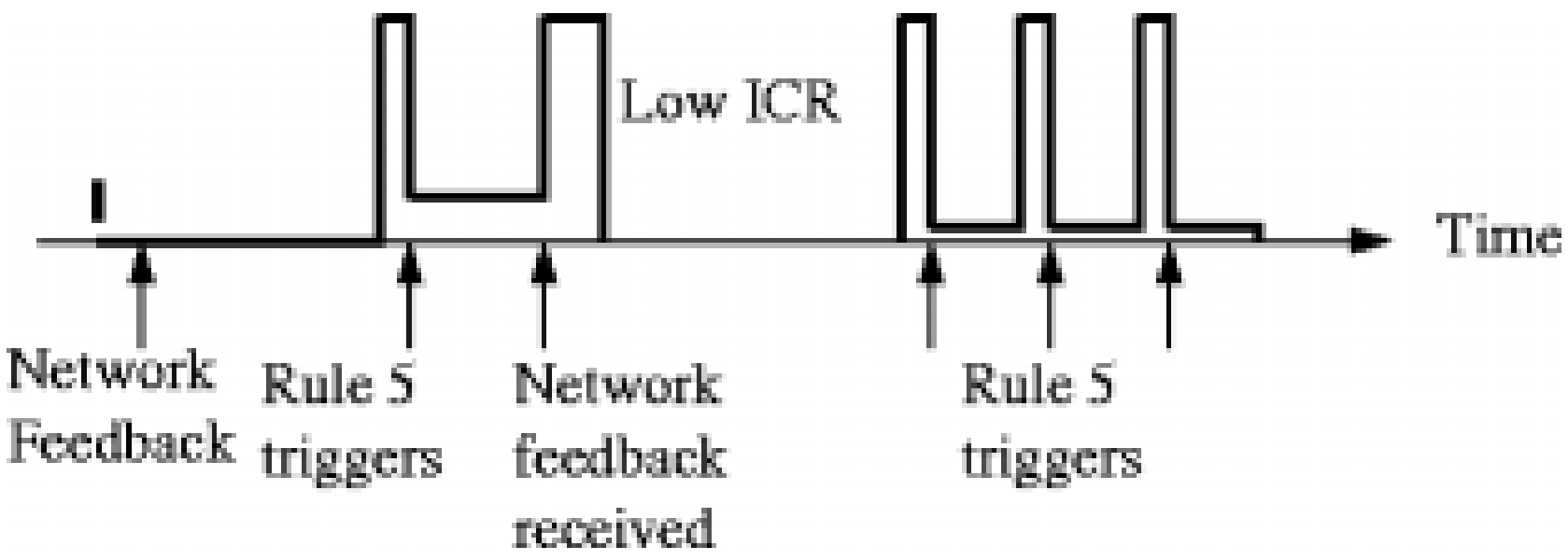

Figure 12: Effect of UILI on Large Bursts